\documentclass{article}[12pts]
\usepackage{setspace}
\usepackage[margin=1in]{geometry}
\usepackage{amsmath}
\usepackage[
backend=biber,
style=numeric-comp,
sorting=none,
maxbibnames=99
]{biblatex}
\renewbibmacro{in:}{}
\usepackage{amssymb}
\usepackage{amsfonts} 
\usepackage{bm}
\bibliography{mybib1.bib}	
\usepackage{listings}
\usepackage{graphicx}
\usepackage[english]{babel}
\usepackage[utf8]{inputenc}
\usepackage{caption}
\usepackage{array,multirow}
\usepackage{subcaption}
\usepackage{xcolor}
\usepackage{hyperref}
\usepackage{float}
\usepackage{tabularx}
\usepackage{adjustbox}
\usepackage[version=3]{mhchem}
\usepackage{siunitx}
\usepackage[capitalize]{cleveref}
\usepackage{authblk}
\title{Bayesian feature selection in joint models with application to a cardiovascular disease cohort study}

\author[1] {Mirajul Islam}
\author[2]{Michael J. Daniels}
\author[3]{Zeynab Aghabazaz}
\author[4]{Juned Siddique}
\affil[1]{PhD Student, Department of Statistics, University of Florida}
\affil[2]{Professor, Department of Statistics, University of Florida}
\affil[3]{Postdoc, Department of Preventive Medicine, Northwestern University Feinberg School of Medicine, Chicago }
\affil[4]{Professor, Department of Preventive Medicine, Northwestern University Feinberg School of Medicine, Chicago}
\date{\today}
\begin{document}
\maketitle
\section{Abstract}
Cardiovascular disease (CVD) cohorts collect data longitudinally to study the association between CVD risk factors and event times. An important area of scientific research is to better understand what features of CVD risk factor trajectories are associated with the disease. We develop methods for
feature selection in joint models where feature selection is viewed as a bi-level variable selection problem with multiple features nested within multiple longitudinal risk factors. We modify a previously proposed Bayesian sparse
group selection (BSGS) prior, which has not been implemented in joint models until now, to better represent prior beliefs when selecting features both at the group level (longitudinal risk factor) and within group (features of a longitudinal risk factor). One of the advantages of our method over the BSGS method is the ability to account for correlation among the features within a risk factor. As a result, it selects important features similarly, but excludes the unimportant features within risk factors more efficiently than BSGS.  We evaluate our prior via simulations and apply our method to data from the Atherosclerosis Risk in Communities (ARIC) study, a population-based, prospective cohort study consisting of over 15,000 men and women aged 45-64, measured at baseline and at six additional times. We evaluate which CVD risk factors and which characteristics of their trajectories (features) are associated with death from CVD. We find that systolic and diastolic blood pressure, glucose, and total cholesterol are important risk factors with different important features associated with CVD death in both men and women.

\section{Introduction}
Longitudinal cardiovascular cohort studies seek to make inferences on the effect of longitudinal risk factor measurements on the time to some event. The Cox model \cite{david1972regression} assumes that the hazard depends only on covariates whose values are constant during follow-up, such as race, sex, and randomized treatment. To handle time dependent exogenous covariates, the extended Cox model \cite{fleming1984nonparametric,andersen1993statistical} has been introduced in the literature. However, this model is not appropriate for endogenous time-dependent covariates like biomarkers.  When primary interest lies in studying the association between such endogenous time-dependent covariates and survival, an alternative modeling framework to the extended Cox model has been introduced in the literature, known as the joint modeling framework for longitudinal and time-to-event data \cite{faucett1996simultaneously,wulfsohn1997joint,tsiatis2004joint}. In these models, the association between time-dependent covariates and events is often characterized by extracting features  from the longitudinal trajectories \cite{rizopoulos2011bayesian,rizopoulos2012joint,rizopoulos2014combining}.\\\\
There are several benefits of jointly modeling survival and covariate data including a reduction in bias in parameter estimation due to covariate measurement error and informative censoring \cite{faucett1996simultaneously}. Many approaches for joint modeling have been discussed in the literature based on single  \cite{wang2001jointly} or double longitudinal outcomes \cite{andrinopoulou2014joint} , recurrent \cite{kim2012joint} or competing event data \cite{andrinopoulou2014joint}, and a nonparametric model for the multiple longitudinal markers \cite{brown2005flexible}. Joint models have also been used to study the association between longitudinal biomarkers and time-to-event in the medical literature, including in cancer \cite{taylor2013real} and HIV studies \cite{buta2015bayesian,dessiso2017bayesian}.\\\\
The choice of a useful model often reduces to the problem of selecting variables to include in the model. There exists various classical approaches to variable selection\cite{hocking1976biometrics,draper1998applied} which are based on sequences of hypothesis tests or estimates of mean squared
error (MSE) \cite{allen1971mean, mallows2000some}. Forward, backward, and stepwise subset selection are commonly used as model/variable selection algorithms. However, these methods have limitations \cite{harrell2017regression} including that variables are either discarded or retained. This discrete process often produces parameter estimates with high variance, and it does not reduce the prediction error of the full model. To overcome these drawbacks, regularization methods \cite{lesaffre2012bayesian, himel2013bayesian} have been proposed. In the Bayesian framework, the model selection problem can be framed in terms of the posterior probabilities of the models (based on carefully chosen priors) instead of searching for a single optimal model \cite{o2009review}. Many methods for Bayesian variable selection have been proposed, including Gibbs Variable Selection (GVS) \cite{dellaportas1982bayesian}, the stochastic search variable selection (SSVS) \cite{george1993variable}, and the Horseshoe prior \cite{carvalho2009handling}; a recent review can be found in \cite{tadesse2021handbook}.\\\\
The literature on variable selection in joint models includes He et al. \cite{he2015simultaneous} who describe a penalized likelihood method with adaptive least absolute shrinkage and selection operator penalty functions for the simultaneous selection of fixed and random effects. In the context of multivariate longitudinal measurements and event time data, Chen and Wang \cite{chen2017variable} developed penalized likelihood methods to select longitudinal features in the survival submodel. For recurrent and terminal events, a variable selection approach was introduced with minimum approximated information criterion in \cite{han2020variable}. Focusing on determining the functional form (feature) that best links a longitudinal process and an event time, a Bayesian shrinkage approach was developed in \cite{andrinopoulou2016bayesian}. To the best of our knowledge, no study has explored feature selection in joint models as a bi-level selection problem with multiple longitudinal processes and multiple features. This is important because in joint models, many feature types are of interest so that we often need to select variables at both the group level and within group. In this paper, instead of taking a traditional Bayesian approach to the group lasso problem \cite{casella2010penalized, raman2009bayesian}, we develop a bi-level variable selection method \cite{xu2015bayesian} in joint models. We propose a novel Bayesian sparse group selection in a joint modeling framework with spike and slab priors to select features both at the group level (longitudinal risk factor) and also within a group (features of longitudinal risk factor). We use the hierarchical spike and slab prior structure of \cite{xu2015bayesian}, which has not been implemented in joint models, and  augment it with a Dirichlet hyperprior (DP) to select variables within each group to better reflect scientific understanding. We will refer to
this as BSGS-Dirichlet (BSGS-D) prior.\\\\
Our work is motivated by data from the ARIC study, a population-based, prospective cohort study of over 15,000 individuals aged 45-64 years at baseline 
\cite{aric1989atherosclerosis}. One of the main objectives of the study was to estimate the effects of the longitudinal risk factors on coronary heart disease (CHD). Baseline information was collected including medical history, physical activity, medication use, and diet during 1987-89. Participants were re-examined 3 times during 1990-1998, then again 3 times during 2011-19. Information collected at each visit included  potential risk factors including body mass index (BMI), systolic blood pressure (SBP), diastolic blood pressure (DBP), glucose level, and total cholesterol (TOTCHL).\\\\
This paper is organized as follows. Section 2 introduces the Bayesian joint model. Section 3 introduces the specification of the  novel Bayesian spike and slab prior. Sections 4 and 5 present a simulation study and the results from the analysis of the ARIC data set,  respectively. Finally, in Section 6, we close with a discussion.
\section{Bayesian joint model}
Let $Y_{ig}(a_{i\ell})$ be the $g$th longitudinal outcome $(g=1,\ldots,G)$ for the $i$th individual $(i=1,\ldots,n)$ at age $a_{i\ell} (\ell=1,2,...,n_i)$. Let $\bm{x}_{ig}$ be the design vector  of the $g$th risk factor for the fixed-effect regression coefficients ($\bm{\beta}_g$) and  $\bm{z_{ig}}$ be corresponding design vector for the random effects \textcolor{red}{($\bm{b}_{ig}$)}. Then, for the $i$th individual, the $g$th longitudinal sub-model is
\begin{eqnarray*}
    y_{ig}(a_{i\ell})&=&\mu_{ig}(a_{i\ell})+\epsilon_{ig}(a_{i\ell}); g=1,2,...,G;~i=1,2,\ldots,n,\\
    \text{where}~\mu_{ig}(a_{i\ell})&=&\bm{x_{ig}}^T(a_{i\ell})\bm{\beta}_g+\bm{z_{ig}}^T(a_{i\ell})\bm{b}_{ig},
\end{eqnarray*}
	where $\bm{\beta_g}=(\beta_{g0},\beta_{g1},\ldots,\beta_{gp})^T,\epsilon_{ig}\sim N(0,\sigma_g^2).$
To account for the correlation among the $G$ longitudinal
outcomes and also correlation within each longitudinal outcome, it is common to assume a multivariate normal distribution for the corresponding random effects for $i$th individual as follows
\begin{equation}\label{eq:2}
	\bm{b}_i = (b_{i1}^T,b_{i2}^T,\ldots,b_{iG}^T)\sim N(\mathbf{0}, \bm{D}).
\end{equation}
For our application, we assume a block diagonal covariance matrix for the random effects in (\ref{eq:2}) where the first block is the covariance among $G$ random intercepts, the second block the covariance among the G random slopes etc. That is,
\begin{equation*}
\bm{D}=Diag(\bm{D}_0,\bm{D}_1,\ldots,\bm{D}_P),~~
\text{where}~~
\bm{D}_p=
\begin{pmatrix}
d_{p11}&\cdots&d_{p1G}\\
d_{p21}&\cdots&d_{p2G}\\
\vdots&\vdots&\vdots\\
d_{pG1}&\cdots&d_{pGG}
\end{pmatrix}
; p=0,1,\ldots,P.
\end{equation*}
Note that the prior development in Section 3 does not rely on this structure.\\
 For the survival outcome, let the observed time $T_i=\text{min}~(T_i^*,C_i)$ where $T_i^*$ and $C_i$ are the true event time and censoring time, respectively. Assuming the risk of the event depends on the $J$ features of the each of the $G$ risk factors and baseline covariates ($w_i$), the survival sub-model is written as
\begin{equation}\label{eq:1}
\lambda_i(t,\bm{\theta_s}) = \lambda_0(t) \exp[\bm{w}_i^T\bm{\gamma}+\sum_{g=1}^{G}\sum_{j=1}^{J}f_{gj}\{\Psi_{ig}(t),\alpha_{gj}\}]
\end{equation}
	where $\Psi_{ig}(t)=\{\mu_{ig}(s),0\leq s\leq t\}$ is the history of the $g$th true unobserved longitudinal process
	up to time point t, $\bm{\theta_s}$ is the parameter vector for the survival outcomes, $\bm{\gamma}$ is the regression coefficients of the baseline covariates, and $\alpha_{gj}$  is a set of parameters that link the longitudinal features with survival outcome. Following  \cite{andrinopoulou2016bayesian}, we specify the baseline hazard function using B-splines,
	$$\log \lambda_0(t) =\gamma_{\lambda_0,0}+\sum_{q=1}^{Q}\gamma_{\lambda_0,q}B_q(t,\bm{k}),$$
	where $(\gamma_{\lambda_0,0},\gamma_{\lambda_0,1},\ldots, \gamma_{\lambda_0,Q})$ are the vector of spline coefficients and $B_q(t,\bm{k})$ denotes the q-th basis function of a B-spline with knots $k_1,\ldots,k_Q$.\\
Aside from priors on $\alpha_g$ which we discuss in Section 3, we assume a Wishart distribution for $\bm{D}$, and diffuse normal priors for the longitudinal regression parameters$(\bm{\beta_g})$ and the baseline survival parameters $\bm{\gamma}$ (see the supplementary materials).
	\subsection{Features}\label{sec:feature}
	Numerous features to link the survival process with longitudinal outcomes have been proposed in the literature \cite{rizopoulos2011bayesian, brown2009assessing}. Here, we consider the following that are most relevant to our ARIC application,
	\begin{eqnarray}
	    \label{eq:3}
		f_{g1}\{\Psi_{ig}(t),\alpha_{g1}\}&=&\alpha_{g1}\mu_{ig}(t)~~~~~~~~~~~~~[\text{value}]\nonumber\\
		f_{g2}\{\Psi_{ig}(t),\alpha_{g2}\}&=&\alpha_{g2}\frac{d\mu_{ig}(t)}{dt}~~~~~~~~~~~[\text{slope}]\nonumber\\
		f_{g3}\{\Psi_{ig}(t),\alpha_{g3}\}&=&\alpha_{g3}\int_{t_0}^t\mu_{ig}(s)ds~~~~~[\text{area}]\nonumber\\
		f_{g4}\{\Psi_{ig}(t),\alpha_{g4}\}&=&\alpha_{g4}I(\mu_{ig}(t)>\Gamma_g) ~~[\text{threshold}]
	\end{eqnarray}
where $f_{g1}(.)$ is the current value of the time-varying risk factor, $f_{g2}(.)$ is the slope of the true trajectory of that risk factor, $f_{g3}(.)$ denotes the area feature, i.e., cumulative effects of the time-varying risk factor  and $f_{g4}(.)$ is the threshold ($\Gamma_g$) feature (whether the current value is above $\Gamma_g$ or not) of the $g$th time-varying risk factor, respectively. It is important to note that the features of the $g$th risk factor are often correlated.  For example, if there is little change over time, the current value and cumulative features will be strongly correlated. Also, the current value and threshold features may also be correlated depending on the value of the threshold. For this reason, there are typically not more than one or two important features for a risk factor. We return to this point in Section 3.1 when we introduce our feature selection prior.  
\section{Specification of the Bayesian spike and slab prior}
As discussed in Section 2, there are different ways to link the longitudinal risk factor and time-to-event. Use of unnecessary features can lead to biased results and issues with collinearity if too many features from the same risk factor are included. Thus, it is important to identify the most appropriate features for each risk factor. To do this, we use a Bayesian spike and slab prior. We first review the BSGS prior proposed in \cite{xu2015bayesian} that we will modify for our setting.\\  
Recall $\bm{\alpha}_g=(\alpha_{g1},\alpha_{g2},\ldots,\alpha_{gJ})^T $ are the parameters that link the features of the $g$th longitudinal risk factor process and the survival process. To tackle sparsity both in risk factor and features within risk factor, we reparametrize the coefficient vectors that link the longitudinal and survival process in (\ref{eq:1}) as follows: 
	$$\bm{\alpha}_g = \bm{V}_g^{\frac{1}{2}}\bm{d_g}, \text{where}~ \bm{V}_g^{\frac{1}{2}}=diag\{\tau_{g1},\tau_{g2},\ldots,\tau_{gJ}\}, \tau_{gj}\geq 0.$$
In the original specification, the same probability was assumed for the group level as well as for the within group level. To select variables at the group level (risk factor), Xu and Ghosh specified the following multivariate spike and slab prior for each $\bm{d_g}$,
	$$\bm{d}_g\overset{ind}{\sim}(1-\pi_{0}) N_{J}(\bm{0}, \bm{I}_{J}) + \pi_{0}\delta_0 (\bm{d_g}),~~~ g = 1,2,\ldots,G,$$
	where  $\delta_0 (\bm{d_g})$ denotes a point mass at $\mathbf{0}\in\mathbb{R}^{J}$. The diagonal elements of $\bm{V}_g^{\frac{1}{2}}$
	control the magnitude of elements of $\bm{\alpha}_g$. Note that when $\tau_{gj} = 0, \alpha_{gj}$ is dropped out of the model even when $d_{gj}\neq 0$.
In order to choose variables (features) within each relevant group (i.e., risk factor), the following spike and slab prior for each $\tau_{gj}$ was introduced
	$$\tau_{gj}\overset{ind}{\sim}(1-\pi_{1})N^+(0,s^2)+\pi_{1}\delta_0(\tau_{gj}),$$
	where $N^+(0,s^2)$ denotes a normal distribution with mean 0 and variance $s^2$ truncated below at 0 and an inverse gamma prior for $s^2$,
 \begin{equation}
  \label{eq:4}
      \frac{1}{s^2}\sim~\text{Gamma}~(1,t)
 \end{equation}
where $t$ is the scale parameter. Conjugate beta hyper-priors were assumed for $\pi_0$  and $\pi_1$,
$$\pi_0\sim \text{Beta}(a, b), \pi_1\sim \text{Beta}(c, d).$$
	
\subsection{BSGS-D specification}
  We introduce several modifications of the original BSGS prior for our setting of risk factor feature selection in joint models. Unlike the original prior, we assume a separate $\pi_g$ for the $g$th risk factor (group) and a separate $\pi_{gj}$ for the $j$th feature within the $g$th risk factor. That is,
$$\bm{d}_g\overset{ind}{\sim}(1-\pi_{g}) N_{G}(\bm{0}, \bm{I}_{J}) + \pi_{g}\delta_0 (d_g), g = 1,2,\ldots,G$$
	$$\tau_{gj}\overset{ind}{\sim}(1-\pi_{gj})N^+(0,s^2)+\pi_{gj}\delta_0(\tau_{gj}).$$
For the risk factor prior selection probabilities, we assume $\pi_g\sim~\text{Beta}~(a,b), g=1,2,\ldots,G$. We want a specification that takes into account that it is unlikely to select more than one or two features for a given risk factor. To accommodate this, we specify a Dirichlet prior to induce correlation among the $\pi_{gj}$ for each $g$ that will implicitly penalize selecting too many features. For the $g$th risk factor, we define the following probabilities,
	\begin{eqnarray*}
q_{gc}=\begin{cases}
\text{the probability of selecting one feature for}~c=1,\ldots,J\\
\text{the probability of selecting two features for}~c=J+1,\ldots,\frac{J(J+1)}{2}\\
\vdots\\
\text{the probability of selecting all J features for}~c=C
\end{cases}
	\end{eqnarray*}
where $C=2^J-1$, since the probability of selecting no features for the $g$th risk factor is \textcolor{red}{implicitly $1$}  when the probability of selecting $g$th risk factor is $0$ (i.e. $\pi_g=0$). The feature selection probabilities ($\pi_{gj}$) are a deterministic function of the $q_{gc}$ and correspond to all possible ways the $j$th feature can be selected. To illustrate, for $J = 3$, the $2^J-1$ possible combinations of feature selection are $(100,010,001,110,101,011,111)$ where $100$ denotes that only the 1st feature is selected, $010$ denotes that only the 2nd feature is selected, and so on. Then  the probability of selecting the 1st, 2nd, and 3rd feature can be computed as $\pi_{g1} = q_{g1} + q_{g4} + q_{g5} + q_{g7},  \pi_{g2}= q_{g2} + q_{g4} + q_{g6} + q_{g7}$, and $\pi_{g3}=q_{g3} + q_{g5} + q_{g6} + q_{g7}$, respectively. More generally, with $J$ features we put a prior on $q_g$ as follows,
$$(q_{g1},q_{g2},\ldots,q_{gC})|a_{g1},a_{g2},\ldots,a_{gJ}\sim~Dirichlet(\bm{a_{g}}),$$
where
$$\bm{a_g}=(\underset{\{J~ times\}}{ a_{g1},\ldots,a_{g1}},\underset{\{\frac{J(J-1)}{2}~ times\}}{a_{g2},\ldots,a_{g2}},\ldots,a_{gJ}).$$
To discourage selecting too many features for $g$th risk factor, we assume $a_{g1}>a_{g2}> a_{g3}>\ldots>a_{g(J-1)}> a_{gJ}>0$. So that, a priori, we are most likely to select one feature per risk factor and least likely to select all features within a risk factor. 
For notational clarity, let $p_g^{(j)}$ be the probability of selecting a feature combination that includes the $j$th feature from the $g$th risk factor. Then, the original $\pi_{gj}$ can be written as
	\begin{eqnarray*}
		\pi_{gj}&=&\sum_{k=0}^{J-1} \binom {J-1}{k}p_g^{(k+1)}.
	\end{eqnarray*} 
The form of the variance and covariance of $\pi_{gj}$ for fixed $g$ can be found in the supplementary materials. To use these priors, we scale the features to have similar magnitudes.\\ 
 We specify the following prior on $a_{gj}$
$$a_{g1},a_{g2},\ldots a_{gJ}\sim \Big\{\prod_{j=1}^{J}~C^+~(0,1)\Big\} I_{\big\{a_{gJ}<a_{g(J-1)}<\cdots<a_{g1}\big\}}.$$
where $C^+(0, 1)$ is a standard half-Cauchy distribution. We use half-Cauchy to allow larger values of the weights of Dirichlet parameters. For large $J$, some combinations of features may be much larger than others. For example, when $J=6$, there are only $\binom J1=6$ one-feature combinations  but  $\binom J3=15$ three-feature combinations. Thus, we also considered an alternative specification of $\bm{a_g}$. This specification (BSGS-D I) differs from the previous specification (BSGS-D) in terms of scaling the weights as follows,
 $$\bm{a_g}=\Bigg(\underset{\{J~ times\}}{ \frac{a_{g1}}{\binom {J}1},\ldots,\frac{a_{g1}}{\binom {J}1}},\underset{\{\frac{J(J-1)}{2}~ times\}}{\frac{a_{g2}}{\binom {J}2},\ldots,\frac{a_{g2}}{\binom {J}2}},\ldots,\frac{a_{gJ}}{\binom {J}J}\Bigg);~~~  \frac{a_{gj}}{\binom J j}>\frac{a_{g(j+1)}}{\binom J {j+1}}).$$
 In this case, the prior on $a_{gj}$ is
$$a_{g1},a_{g2},\ldots a_{gJ}\sim \Big\{\prod_{j=1}^{J}~C^+~(0,1)\Big\} I_{\big\{\frac{a_{gJ}}{\binom {J}J}<\frac{a_{g(J-1)}}{\binom {J}{J-1}}<\cdots<\frac{a_{g1}}{\binom {J}1}\big\}}.$$
The covariance in terms of Dirichlet parameters can be expressed as
	\begin{eqnarray} \label{eq:cov}
Cov(\pi_{gi},\pi_{gk}) &=&C\Bigg(-a_{g1}^2-2\sum_{i=0}^{J-2}\sum_{j=i+1}^{J-1}\binom {J-1}i\binom{J-1}j a_{g(i+1)}a_{g(j+1)}-\sum_{j=0}^{J-3}\Bigg[{\binom{J-1}{j+1}}^2-\binom{J-2}{j}\Bigg]a_{g(j+2)}^2\nonumber\\
 &&+\sum_{j=0}^{J-2}\binom{J-2}{j}a_{g(j+2)}(a_{gt}-a_{g(j+2)})\Bigg)
	\end{eqnarray}
 where $C=\frac{1}{a_{gt}^2(a_{gt}+1)}>0$ with $a_{gt}=\sum_{j=1}^J\binom Jj a_{gj}$. We show the covariance between the selection probabilities of any two features within a risk factor is negative for BSGS-D prior for $J \in \{2,3,4\}$.  For $J=2$ and $J=3$,
 \begin{eqnarray*}
   Cov(\pi_{gj},\pi_{gk})=\begin{cases}{-Ca_{g1}^2<0;~\text{for}~J=2}\\
   {-C\Big(a_{g1}^2+a_{g2}^2+a_{g1}(a_{g2}-a_{g3})\Big)<0;~\text{for}~J=3}.
   \end{cases}
 \end{eqnarray*}
Similarly, for $J=4$,
\begin{eqnarray*}
    Cov(\pi_{gj},\pi_{gk})&=&-C\Big(a_{g1}^2+3a_{g2}^2+a_{g3}^2+2a_{g1}(a_{g2}-a_{g3})+a_{g2}(2a_{g3}-a_{g4})-2a_{g1}a_{g4}\Big); j\neq k\\
    &<&C\Big(-a_{g1}^2-3a_{g2}^2-a_{g3}^2-2a_{g1}(a_{g2}-a_{g3})-2a_{g2}(a_{g3}-a_{g4})-a_{g2}a_{g4}+2a_{g1}a_{g4}+{(a_{g1}-a_{g4})}^2\Big)\\
    &=&-C\Big(3a_{g2}^2+(a_{g3}^2-a_{g4}^2)+2a_{g1}(a_{g2}-a_{g3})+2a_{g2}(a_{g3}-a_{g4})+a_{g2}a_{g4}\Big)\\
    &<&0.
\end{eqnarray*}
We conjecture that the covariance will be negative for $J \in \{5,6,\ldots\}$ but we have been unable to prove it for general $J$. 

 For both priors, we estimate $t$ in (\ref{eq:4}) first using posterior samples,
	$$\hat{t}=\frac{1}{E[\frac{1}{s^2}|data]}$$
where the initial rate parameter is set to $1$ for the gamma distribution in (\ref{eq:4}). To use these priors in practice we standardize the features as in \cite{andrinopoulou2016bayesian}.
\subsection{Posterior Sampling}
The posterior distribution of the fixed effect, random effect, and covariance parameters can be written as 

$$p(\bm{\theta},\bm{b}| y_{i}, T_i,\Delta_i)\propto\prod_{g=1}^G\prod_{\ell=1}^{n_i}p(y_{i\ell g}|b_{ig},\theta_{y_g})p(T_i,\Delta_i|\Psi_{ig}(T_i),\theta_s)p(b_{ig}|\theta_{y_{g}})p (\theta_{y_g})p(\theta_s),$$
where $\bm{\theta}=(\bm{\theta}_s^T,\bm{\theta}_{y_g}^T,\bm{D})~\text{and}~\bm{\theta}_{y_g}=(\bm{\beta}_g^T,\sigma_g),\bm{\theta}_s=(\bm{\gamma}^T,\bm{\alpha_1}^T,\bm{\alpha_2}^T,\ldots,\bm{\alpha_G}^T)$.
 We sample from the posterior using JAGS \cite{plummer2003jags} in R studio \cite{team2018rstudio}.
	\section{Simulation study}
	\subsection{Design}
	We conducted a simulation study to assess the performance of the BSGS-D prior. We simulated data similar to ARIC, with 800 subjects including some censored individuals (also considered a smaller (400) and a larger (1600) number of subjects) and a maximum of 5 follow-up visits. The censoring rate varied by the simulation scenario but ranged from $5\%$ to $50\%$.
 We assumed a linear mixed-effects model for each of the three longitudinal risk factors,
	$$y_{ig}(a_{i\ell})=\beta_{g0}+\sum_{k=1}^2\beta_{k,g}P_k(a_{i\ell})+b_{i0,g}+\sum_{k=1}^2b_{ik,g}P_k(a_{i\ell}) +\epsilon_{ig}(a_{i\ell}),$$
	where $P_1(a_{i\ell})=(\frac{2a_{i\ell}}{a_{max}}-1); P_2(a_{i\ell})=\frac{1}{2}( 3(\frac{2a_{i\ell}}{a_{max}}-1)^2-1); g=1,2,3;~i=1,2,\ldots,800,\epsilon_{ig}\sim N(0,\sigma_g^2)~\text{and}~\bm{b_i}\sim N(\bm{0},\bm{D})$ with $\bm{D}$ assumed to be a block diagonal matrix as in  Section 2. We used  a second order shifted Legendre orthogonal polynomial for a nonlinear effect for age for both the fixed and the random effects. The baseline (centered) age was simulated uniformly from 0 to 19. Following ARIC, the subsequent visits are taken at three-year intervals. Race was simulated as a Bernoulli $(0.5)$.  For the survival model, we used a proportional hazards model assuming a B-splines baseline hazard and including all four possible features for each risk factor,
	\begin{eqnarray}\label{eq:surv}
				\lambda_i(t)& =& \lambda_0(t) \exp\Big[\gamma{Race}_i+\sum_{g=1}^{G}\Big(\alpha_{g1}\mu_{ig}(t)+\alpha_{g2}\frac{d\mu_{ig}(t)}{dt}+\alpha_{g3}\int_{t_0}^t\mu_{ig}(s)ds+\alpha_{g4}I\big(\mu_{ig}(t)>\Gamma_g\big)\Big)\Big].
		\end{eqnarray}
 We considered the following scenarios:
 \begin{itemize}
    \item \textbf{Scenario I}: one important risk factor with one feature corresponding to be current value of the risk factor  
       $$\lambda_i(t)=\lambda_0(t)\exp\big[\gamma {\text{Race}}_i+\alpha_{21}\mu_{i2}(t)\big].$$
       \item \textbf{Scenario II}: all three important risk factors, each with one, but different important features  corresponding to be area, current value, and slope   
       $$\lambda_i(t)=\lambda_0(t)\exp\big[\gamma {\text{Race}}_i+\alpha_{13}\int_{t_0}^t\mu_{i1}(s)ds+\alpha_{21}\mu_{i2}(t)+\alpha_{32}\frac{d}{dt}\mu_{i3}(t)\big].$$
     \item \textbf{Scenario III}: one important risk factor with two important features  corresponding to be current value and area
      $$\lambda_i(t)=\lambda_0(t)\exp\big[\gamma {\text{Race}}_i+\alpha_{11}\mu_{i1}(t)+\alpha_{13}\int_{t_0}^t\mu_{i1}(s)ds\big].$$
        \item \textbf{Scenario IV}: two important risk factors with two important features
       $$\lambda_i(t)=\lambda_0(t)\exp\big[\gamma {\text{Race}}_i+\alpha_{11}\mu_{i1}(t)+\alpha_{13}\int_{t_0}^t\mu_{i1}(s)ds+\alpha_{21}\mu_{i2}(t)+\alpha_{23}\int_{t_0}^t\mu_{i2}(s)ds\big].$$
\end{itemize}
It is unlikely that more than 2 features within a risk factor are important. We chose the first two scenarios to evaluate the performance of our method when there is only one important feature case within a risk factor. We also consider scenarios III and IV to compare the performances when more than one feature within a risk factor is important. \\
We fit a joint model using the linear mixed model above Cox models as above but using all risk factors and features to compare the performance of the BSGS-D prior, and a 'standard' spike and slab (SS) prior to the original BSGS prior. The SS prior is defined following \cite{bai2021spike} as

$$\alpha_{gj}\overset{ind}{\sim}(1-\pi_{gj})N(0,1/\tau^2)+\pi_{gj}\delta_0(\alpha_{gj})$$
$$\pi_{gj}\sim Beta(1,10), \tau^2\sim Gamma(1,1).$$ 
For each scenario, the posterior distribution was sampled for 100 replicated datasets and  the priors were compared in terms of average percentages of feature selection, bias, and mean squared error (MSE) of the feature parameters ($\alpha_{gj}$).

\subsection{Results}
 Table \ref{1R1F_3R3F1} shows the comparison between the BSGS-D prior, the original BSGS prior and SS prior for $n=800$. For scenario I, where only the second risk factor was important with one important feature, this feature is selected 88\% of the time by BSGS-D prior, 92\% by BSGS prior and 60\% by SS prior. However, the original BSGS prior notably overselects the non-important features within the second risk factor and SS prior underselects the important feature. For scenario II where all three risk factors were important with one important feature for each, the BSGS-D method again selects the important features at a similar rate to BSGS and better rate to SS prior but better excludes the unimportant features compared to BSGS. For the larger sample size ($n=1600$) shown in  Table \ref{1R1F_3R3F2} for Scenarios I and II, we see the improvement for all three priors in terms of selection but  BSGS prior still overselects the non-important features within the second risk factor. We considered the two features for the first risk factor, and for both first and second risk factor as important features for the survival outcome in Scenario III and Scenario IV, respectively. These two scenarios are presented in Table \ref{1R2F_2R2F1} for sample 800 and Table \ref{1R2F_2R2F2} for sample 1600. They again show that a (slightly) higher percentage of important features were selected using BSGS. However, the BSGS-D prior again selects unimportant features within a risk factor with a much smaller probability and better selects important features within a risk factor compared to SS prior. For example, in scenario III for $n=800$, value and area features of the first risk factor were selected approximately 100\% of the time by the BSGS-D prior and BSGS prior. However, the BSGS prior also selected the unimportant slope and threshold features 47\% and 48\% of the time while the BSGS-D prior only selected them 29\% and 25\%, respectively. SS prior underselects the important features compared to other two priors. Overall, BSGS-D I performs similarly to BSGS-D.\\\\
For all the scenarios except Scenario I, the bias of the coefficients of the true important features for the BSGS-D prior are smaller than or equal  to the bias for BSGS (see Tables S.1 to S.12 of the supplementary materials). The MSEs of the coefficients of the important features are similar among all three priors except SS prior. SS prior produces larger bias and MSE compared to other three priors for sample sizes 800 and 1600.\\\\
All four priors produce biased (attenuated towards zero) results when more than one feature within a risk factor is important. This bias can be reduced by fitting the model again with only the important features.\\\\
 Results including average percentage of feature selection for the smaller sample size ($n=400$) are shown in the supplementary materials (see Tables S.13 and S.14). The results of different sample sizes show that selection probabilities increase for true risk factors, and true features and decrease for unimportant risk factors, and unimportant features as sample size increases as expected. 

\section{Analysis of ARIC data}
In this section, we present the analysis of the ARIC data. The goal is to identify important risk factors as well as features within important risk factors that are associated with CHD death. We separately analyze male and female participants. We used mixed-effect longitudinal submodels as in the simulation for the risk factor with quadratic Legendre orthogonal polynomials for age in both the fixed-effects and random-effects parts of the models as the profiles were non-linear.  We also included race and education in the submodel.\\
For the survival submodels, we assumed proportional hazards models, modeling the baseline hazard function with a B-spline basis (as in simulation). We considered four features for each of the five longitudinal risk factors (BMI, SBP, DBP, GLUCOSE, and TOTCHL): the value, the threshold, the slope, and the area under the curve as defined in Section \ref{sec:feature} . Following \cite{grotto2006prevalence}, the threshold values ($\Gamma_g$) were specified as 30 (kg/m2) for BMI, 120 (mmHg) for SBP, 80 (mmHg) for DBP, 126 (mg/dL) for GLUCOSE and 230 (mg/dL) for TOTCHL. The survival sub-model takes the form
\begin{eqnarray*}
				\lambda_i(t)& =& \lambda_0(t) \exp\Big[\gamma{\text{Race}}_i+\sum_{g=1}^{5}\Big(\alpha_{g1}\mu_{ig}(t)+
		\alpha_{g2}\frac{d\mu_{ig}(t)}{dt}+\alpha_{g3}\int_{t_0}^t\mu_{ig}(s)ds+\alpha_{g4}I\big(\mu_{ig}(t)>\Gamma_g\big)\Big)\Big].
		\end{eqnarray*}
Table \ref{ARIC_res} shows the posterior probabilities of risk factor and feature selection. SBP, DBP, GLUCOSE, and TOTCHL are important risk factors for CHD death for both males and females (as expected), as these risk factors were selected more than 90\% of the time by both BSGS-D and BSGS priors. BMI is also an important risk factor for CHD death but only among females. Both slope and area features of BMI are important for female by all three priors but are selected with only 56\% and 53\% of the time by SS prior.  For SBP, the area and threshold features are important among males whereas the value feature is the most important among females. The important features of DBP are area for male, and value and threshold for female.  The most important feature of GLUCOSE for males and females is the threshold feature. While the value feature of GLUCOSE is important by BSGS prior (selection percentage 83\%), this is not important according to BSGS-D prior (selection percentage 56\%), and SS prior (selection percentage 10\%).   No individual feature of TOTCHL shows strong evidence as being important. The area feature of TOTCHL has two of the three largest effects on CHD death among males and females (\ref{ARIC_res_det}) and the two highest selection probabilities.\\\\
BMI is an important risk factor for males according to BSGS prior, but it is not important under BSGS-D and SS priors. We also note that original BSGS prior likely overselects potentially unimportant features based on the results of the simulation study. For example, the potentially unimportant features, value, slope, and threshold of DBP among men were selected 55\%, 54\%, and 69\%, respectively by BSGS prior (vs. 24\%, 12\%, and 35\% for the BSGS-D prior). The SS prior underselects the important features compared to other two priors. For example, the selection percentages for threshold feature of SBP among males are 89\% by BSGS-D prior, 96\% by BSGS prior, respectively but only 49\% by SS prior. Similarly, area feature of DBP is selected only 66\% of the time  by SS prior compared to 79\% by BSGS-D, and 90\% by BSGS priors. \\\\

The posterior mean features and 95\% CI for the BSGS-D prior, original BSGS prior, and SS prior are given in Table \ref{ARIC_res_det}. Here, the potential over-selection of unimportant features in an important risk factor can be seen by the larger posterior mean of coefficients that are potentially unimportant (lower probability of selection) between BSGS and BSGS-D (see e.g., the slope coefficient for TOTCHL). Similarly, the potential under-selection of important features by SS prior can be seen by the smaller posterior mean of coefficients that are potentially important by all three priors (e.g., the area coefficient of DBP and threshold coefficient of SBP for males).  	
 
\section{Discussion}
One of the main objectives of the ARIC study was to investigate the association between longitudinal risk factors and coronary heart disease (CHD) death \cite{fisher1999time}. In this paper, we focused on finding the functional form of risk factor trajectories and investigated which features of the risk factors were associated with CHD death.  To identify the most appropriate feature(s) from each risk factor,  we introduced a Bayesian sparse group selection with spike and slab Prior (BSGS) augmented with a Dirichlet prior. The new proposed prior, BSGS-D is constructed to account for the fact that it is unlikely to select multiple features for a given risk factor. For ARIC data, we have found that all the considered risk factors except BMI had a significant effect on CHD death for males and females. However, the important features for each risk factor varied by sex.\\\\
From the comparison between BSGS-D prior and original BSGS prior, we came to a conclusion after simulations that our prior appears to have better performance to deselect the unimportant features by accounting for correlation among the features within a risk factor. In particular, our prior selects important features similarly to the BSGS-prior but deselects the unimportant features within a risk factor with (sometimes much) higher probabilities. The BSGS-D prior has no computational cost compared to the original BSGS prior.  The SS prior under-selects the important features and produces more biased results than BSGS-D, and original BSGS priors.  So either the modified BSGS prior, BSGS-D, or the original BSGS prior would be preferred to a standard SS prior for feature selection in joint models.\\\\
We assumed a multivariate normal distribution for the random effects. The model could be extended to a semiparametric Bayesian joint model by assuming a Dirichlet process mixture (DPM) for the distribution of the random effects \cite{rizopoulos2011bayesian}. One could also replace the default priors for the coefficients on baseline covariates in the longitudinal and survival submodels with spike and slab priors. The framework could also be extended to address competing risks and estimating a data dependent threshold parameter in the threshold feature.   

\hspace{-0.5cm}\textbf{Acknowledgements}\\
All authors were partially supported by NIH R01 HL 158963. ARIC data are available through the NHLBI BIOLINCC data repository.

\medskip
\printbibliography
 \begin{table}[H]
   \centering
		\caption{Average percentage of feature selection for $n=800$~(Scenarios I \& II: only one important feature within a risk factor). Bold text indicates true important risk factors and features}
   \label{1R1F_3R3F1}
		\begin{tabular}{c|cccccc}
			\hline
			Scenario&Risk factor&Feature&\multicolumn{4}{c}{Percentage of selection (MC SE)}\\
			&&&BSGS-D&BSGS&BSGS-D I&SS\\\hline
			\multirow{12}{*}{Scenario-I}&First&&27 (1.1)&34 (1.0)&27 (1.0)&-\\
                &&value&6 (0.8)&8 (0.9)&6 (0.7)&2 (0.5)\\
			&&slope&7 (1.0)&8 (0.7)&7 (0.9)&2 (0.4)\\
			&&area&6 (0.5)&8 (0.7)&6 (0.4)&2 (0.2)\\
			&&threshold&8 (0.7)&9 (0.7)&8 (0.6)&3 (0.4)\\\cline{2-7}
        &\textbf{Second}&&100 (0.0)&100 (0.0)&100 (0.0)&-\\
		&&\textbf{value}&88 (2.0)&92 (1.6)&88 (2.0)&60 (3.7)\\
			&&slope&23 (1.5)&33 (1.4)&23 (1.1)& 3 (0.7)\\
			&&area&30 (1.4)&38 (1.4)&30 (1.4)& 6 (1.0)\\
			&&threshold&44 (1.9)&50 (1.7)&46 (2.0)& 47 (3.6)\\\cline{2-7}
			&Third&&29 (1.1)&37 (1.2)&31 (1.3)&-\\
   &&value&7 (0.9)&10 (1.1)&8 (0.9)&2 (0.3)\\
			&&slope&7 (0.8)&10 (1.0)&8 (1.0)&2 (0.5)\\
			&&area&7 (0.6)&9 (0.9)&7 (0.8)&2 (0.2)\\
			&&threshold&11 (0.8)&13 (1.1)&12 (1.0)& 6 (0.8)\\\hline
			\multirow{12}{*}{Scenario-II}&\textbf{First}&&100 (0.0)&100 (0.0)&100 (0.0) &-\\
   &&value&34 (1.7)&57 (1.5)&34 (1.6)&13 (2.4)\\
			&&slope&26 (1.8)&50 (1.6)&24 (1.5)&8 (1.8)\\
			&&\textbf{area}&93 (1.3)&96 (0.7)&93 (1.3)&89 (2.5)\\
			&&threshold&33 (1.1)&59 (1.2)&33 (1.0)&7 (1.3)\\\cline{2-7}
			&\textbf{Second}&&100 (0.0)&100 (0.0)&100 (0.0)&-\\
   &&\textbf{value}&94 (1.5)&98 (0.6)&93 (1.6)& 74 (3.4)\\
			&&slope&31 (2.0)&56 (1.9)&33 (2.2)&25 (3.2)\\
			&&area&40 (2.0)&62 (1.8)&42 (2.1)& 15 (2.4)\\
			&&threshold&41 (2.0)&64 (1.5)&45 (1.8)&36 (3.4)\\\cline{2-7}
			&\textbf{Third}&&95 (1.4)&94 (1.4)&93 (1.8)&-\\
   &&value&22 (1.3)&46 (1.5)&21 (1.1)&6 (1.1)\\
			&&\textbf{slope}&91 (2.0)&90 (1.9)&89 (2.4)&51 (3.6)\\
			&&area&24 (0.9)&49 (1.5)&22 (1.0)&23 (3.0)\\
			&&threshold&34 (1.3)&57 (1.4)&32 (1.2)&12 (1.7)\\
			\hline
		\end{tabular}
	\end{table}
 \begin{table}[H]
   \centering
		\caption{Average percentage of feature selection for $n=1600$~(Scenarios I \& II: only one important feature within a risk factor). Bold text indicates true important risk factors and features}
   \label{1R1F_3R3F2}
		\begin{tabular}{c|cccccc}
			\hline
			Scenario&Risk factor&Feature&\multicolumn{4}{c}{Percentage of selection (MC SE)}\\
			&&&BSGS-D&BSGS&BSGS-D I&SS\\\hline
			\multirow{12}{*}{Scenario-I}&First&&24 (0.8)&32 (0.9)&23 (0.9)&- \\
                &&value&4 (0.4)&5 (0.4)&4 (0.3)&1 (0.1)\\
			&&slope&5 (0.6)&6 (0.7)&5 (0.8)&1 (0.5)\\
			&&area&5 (0.4)&6 (0.5)&5 (0.3)&1 (0.2)\\
			&&threshold&6 (0.4)&7 (0.5)&6 (0.5)&3 (0.3)\\\cline{2-7}
        &\textbf{Second}&&100 (0.0)&100 (0.0)&100 (0.0)&-\\
		&&\textbf{value}&95 (1.5)&95 (1.3)&96 (1.2)&61 (3.6)\\
			&&slope&23 (1.5)&31 (1.7)&21 (1.6) &3 (1.1)\\
			&&area&26 (1.3)&36 (1.6)&25 (1.2)& 6 (1.2)\\
			&&threshold&39 (2.0)&46 (2.0)&37 (1.6)&49 (3.7)\\\cline{2-7}
			&Third&&28 (1.4)&35 (1.1)&27 (1.3)&-\\
   &&value&8 (1.3)&8 (1.1)&7 (1.0)&3 (0.7)\\
			&&slope&6 (0.7)&7 (0.9)&5 (0.7)&1 (0.1)\\
			&&area&6 (0.6)&8 (0.9)&6 (0.7)& 2 (0.4)\\
			&&threshold&9 (0.7)&10 (0.8)&8 (0.6)& 4 (0.3) \\\hline
			\multirow{12}{*}{Scenario-II}&\textbf{First}&&100 (0.0)&100 (0.0)&100 (0.0)&-\\
   &&value&24 (1.3)&46 (1.6)&25 (1.2)&8 (1.6)\\
			&&slope&21 (1.7)&41 (2.0)&19 (1.4)&8 (1.9)\\
			&&\textbf{area}&98 (1.0)&98 (0.4)&98 (0.8)&95 (1.6)\\
			&&threshold&29 (1.3)&49 (1.4)&29 (1.1)&5 (1.0)\\\cline{2-7}
			&\textbf{Second}&&100 (0.0)&100 (0.0)&100 (0.0)&-\\
   &&\textbf{value}&93 (1.7)&98 (1.1)&94 (1.7)&76 (2.8)\\
			&&slope&28 (1.9)&45 (2.1)&26 (1.8)&31 (3.2)\\
			&&area&39 (2.3)&53 (1.8)&40 (2.2)&21 (2.6)\\
			&&threshold&48 (2.3)&58 (1.9)&48 (2.6)&49 (3.1)\\\cline{2-7}
			&\textbf{Third}&&100 (0.0)&99 (0.4)&99 (0.7)&-\\
   &&value&20 (1.1)&40 (1.3)&20 (1.2)&9 (1.7)\\
			&&\textbf{slope}&100 (0)&99 (0.6)&98 (1.0)&59 (3.4)\\
			&&area&20 (1.0)&40 (1.2)&20 (0.7)&21 (2.5)\\
			&&threshold&31 (0.9)&51 (1.2)&32 (0.9)&11 (1.2)\\
			\hline
		\end{tabular}
	\end{table}

\begin{table}[H]
   \centering
		\caption{Average percentage of feature selection for $n=800$~(Scenarios III \& IV: more than one important feature within a risk factor). Bold text indicates true important risk factors and features}
   \label{1R2F_2R2F1}
		\begin{tabular}{c|cccccc}
			\hline
			Scenario&Risk factor&Feature&\multicolumn{4}{c}{Percentage of selection (MC SE)}\\
			&&&BSGS-D&BSGS&BSGS-D I&SS\\\hline
			\multirow{12}{*}{Scenario-III}&\textbf{First}&&100 (0.0)&100 (0.0)&100 (0.0)&-\\
   &&\textbf{value}&100 (0.0)&100 (0.0)&100 (0.0)&100 (0.0)\\
			&&slope&29 (2.4)&47 (2.3)&27 (2.5)&8 (1.8)\\
			&&\textbf{area}&100 (0.0)&100 (0.0)&100 (0.0)&98 (1.0)\\
			&&threshold&25 (1.0)&48 (1.3)&27 (1.1)&5 (0.7)\\\cline{2-7}
			&Second&&22 (1.4)&19 (1.6)&24 (1.3)&-\\
   &&value&4 (0.9)&5 (1.0)&5 (1.0)&4 (1.2)\\
			&&slope&5 (1.1)&4 (1.0)&4 (0.9)&5 (1.5)\\
			&&area&4 (0.5)&5 (0.9)&4 (0.5)&1 (0.4)\\
			&&threshold&6 (0.8)&6 (1.3)&8 (1.2)&6 (1.3)\\\cline{2-7}
			&Third&&21 (0.9)&18 (1.1)&22 (0.9)&-\\
   &&value&3 (0.5)&4 (0.7)&3 (0.4)&1 (0.2)\\
   &&slope&3 (0.4)&5 (0.9)&4 (0.8)&5 (1.5)\\
			&&area&3 (0.5)&5 (0.8)&4 (0.4)&3 (1.2)\\
			&&threshold&7 (0.7)&7 (0.9)&7 (0.6)&4 (0.5)\\\hline
   \multirow{12}{*}{Scenario-IV}&\textbf{First}&&99 (0.3)&99 (0.4)&99 (0.4)&-\\
   &&\textbf{value}&88 (2.2)&95 (1.0)&89 (2.2)&83 (2.8)\\
			&&slope&46 (3.1)&73 (2.2)&50 (2.9)&26 (3.1)\\
			&&\textbf{area}&90 (2.0)&97 (0.9)&90 (2.0)&85 (2.6)\\
			&&threshold&30 (1.2)&67 (1.2)&30 (1.2)& 5 (0.8)\\\cline{2-7}
			&\textbf{Second}&&99 (0.6)&100 (0.3)&99 (0.6)&-\\
   &&\textbf{value}&92 (1.7)&98 (0.7)&92 (1.6)&68 (3.3)\\
			&&slope&40 (2.9)&67 (2.0)&37 (2.7)&31 (3.5)\\
			&&\textbf{area}&83 (2.1)&91 (1.2)&84 (2.2)&54 (3.6)\\
			&&threshold&42 (2.4)&74 (1.5)&42 (2.3)&35 (3.5)\\\cline{2-7}
			&Third&&36 (2.1)&48 (2.7)&36 (1.9)&-\\
   &&value&15 (2.0)&34 (2.8)&15 (1.9)&21 (2.7)\\
			&&slope&11 (1.6)&30 (2.6)&10 (1.4)&13 (2.7)\\
			&&area&12 (1.3)&32 (2.5)&12 (1.5)&21 (2.7)\\
			&&threshold&13 (1.2)&34 (2.4)&13 (1.3)&8 (1.3)\\\hline
		\end{tabular}
	\end{table}
\begin{table}[H]
   \centering
		\caption{Average percentage of feature selection for $n=1600$~(Scenarios III \& IV: more than one important feature within a risk factor). Bold text indicates true important risk factors and features}
   \label{1R2F_2R2F2}
		\begin{tabular}{c|cccccc}
			\hline
			Scenario&Risk factor&Feature&\multicolumn{4}{c}{Percentage of selection (MC SE)}\\
			&&&BSGS-D&BSGS&BSGS-D I&SS\\\hline
			\multirow{12}{*}{Scenario-III}&\textbf{First}&&100 (0.0)&100 (0.0)&100 (0.0)&-\\
   &&\textbf{value}&100 (0.0)&100 (0.0)&100 (0.0)&100 (0.0)\\
			&&slope&31 (2.6)&48 (2.7)&31 (2.8)&12 (2.4)\\
			&&\textbf{area}&99 (0.5)&99 (0.3)&100 (0.0)&100 (0.0)\\
			&&threshold&21 (1.1)&46 (1.4)&24 (1.4)&3 (0.5)\\\cline{2-7}
			&Second&&31 (2.4)&30 (2.1)&30 (2.3)&-\\
   &&value&11 (2.0)&13 (1.7)&12 (1.9)&14 (2.3)\\
			&&slope&5 (0.8)&10 (1.3)&6 (1.1)& 5 (1.4)\\
			&&area&6 (1.1)&10 (1.2)&6 (1.0)&2 (0.7)\\
			&&threshold&18 (2.7)&18 (2.2)&16 (2.4)&19 (2.6)\\\cline{2-7}
			&Third&&22 (1.0)&22 (1.7)&22 (1.2)&-\\
   &&value&3 (0.6)&6 (0.8)&3 (0.4)&1 (0.5)\\
   &&slope&4 (0.8)&6 (1.2)&3 (0.5)&5 (1.3)\\
			&&area&3 (0.6)&6 (1.0)&4 (0.6)&1 (0.5)\\
			&&threshold&7 (0.8)&9 (1.3)&7 (0.9)&5 (0.9)\\\hline
 \multirow{12}{*}{Scenario-IV}&\textbf{First}&&100 (0.0)&100 (0.0)&100 (0.0)&-\\
   &&\textbf{value}&99 (1.0)&99 (0.5)&98 (0.8)&94 (1.6)\\
			&&slope&44 (3.0)&74 (2.2)&45 (2.6)&24 (3.0)\\
			&&\textbf{area}&99 (0.6)&99 (0.4)&99 (0.6)&96 (1.3)\\
			&&threshold&26 (1.0)&66 (1.2)&25 (0.9)&3 (0.4)\\\cline{2-7}
			&\textbf{Second}&&100 (0.0)&100 (0.0)&99 (2.8)&-\\
   &&\textbf{value}&95 (1.5)&99 (0.4)&97 (1.1)&73 (3.2)\\
			&&slope&43 (3.1)&65 (2.1)&41 (2.7)&38 (3.0)\\
			&&\textbf{area}&86 (2.1)&98 (0.6)&90 (2.0)&66 (3.5)\\
			&&threshold&51 (2.8)&78 (1.7)&52 (3.0)&55 (3.7)\\\cline{2-6}
			&Third&&51 (2.7)&57 (3.1)&46 (2.7)&-\\
                &&value&32 (3.1)&46 (3.3)&28 (3.1)&45 (3.8)\\
			&&slope&17 (2.0)&36 (2.7)&14 (1.7)&27 (3.1)\\
			&&area&22 (2.4)&43 (3.2)&21 (2.7)&42 (3.7)\\
			&&threshold&19 (1.4)&42 (2.7)&17 (1.4)&9 (1.4)\\\hline
		\end{tabular}
	\end{table} 
\begin{table}[H]
   \centering
		\caption{Posterior probabilities of risk factors and features within a risk factor in ARIC for male and female}
    \label{ARIC_res}
		\begin{tabular}{ccccc|ccc}
			\hline
			Risk factor&Feature&\multicolumn{6}{c}{Percentage of selection}\\\hline
   &&\multicolumn{3}{c|}{Male}&\multicolumn{3}{c}{Female}\\
   &&BSGS-D&BSGS&SS&BSGS-D&BSGS&SS\\
   \hline
   BMI&&47&82&-&99&100&-\\
	 &value&12&44&2&31&65&10\\
      &slope&28&63&4&80&89&56\\
      &area&22&57&7&71&89&53\\
      &threshold&13&46&1&30&64&6\\
			\hline
   SBP&&100&100&-&100&100&-\\
	 &value&51&68&11&100&100&100\\
      &slope&12&36&1&12&48&0\\
      &area&86&96&94&23&68&2\\
      &threshold&89&96&49&56&82&16\\
			\hline
    DBP&&91&99&-&100&100&-\\
	 &value&24&55&3&93&91&99\\
      &slope&12&54&26&22&62&1\\
      &area&79&90&66&39&79&3\\
      &threshold&35&69&20&91&95&83\\
			\hline
   GLUCOSE&&100&100&-&100&100&-\\
	 &value&23&45&2&56&83&10\\
      &slope&52&45&69&14&47&1\\
      &area&47&69&14&69&81&92\\
      &threshold&100&100&100&98&99&87\\
			\hline
   TOTCHL&&100&100&-&99&100&-\\
	 &value&40&63&19&38&67&64\\
      &slope&26&66&2&46&74&1\\
      &area&69&63&73&63&79&35\\
      &threshold&47&73&14&31&68&6\\
			\hline
		\end{tabular}
	\end{table}

 \begin{table}[H]
\centering
		\caption{Posterior distribution summaries including posterior mean and 95\% CI for the association parameters for features in ARIC under BSGS, BSGS-D and SS priors }
    \label{ARIC_res_det}
    \begin{adjustbox}{width=18cm }
		\begin{tabular}{cc|ccccccccc|ccccccccc}
			\hline
			Risk factor&Feature&\multicolumn{18}{c}{Posterior summary}\\\hline
   &&\multicolumn{9}{c|}{Male}&\multicolumn{9}{c}{Female}\\
   &&\multicolumn{3}{c}{BSGS-D}&\multicolumn{3}{c}{BSGS}&\multicolumn{3}{c|}{SS}&\multicolumn{3}{c}{BSGS-D}&\multicolumn{3}{c}{BSGS}&\multicolumn{3}{c}{SS}\\
&&Mean&2.5\%&97.5\%&Mean&2.5\%&97.5\%&Mean&2.5\%&97.5\%&Mean&2.5\%&97.5\%&Mean&2.5\%&97.5\%&Mean&2.5\%&97.5\%\\
   \hline
	 BMI&value&0.004&-0.025&0.100&0.013&-0.152&0.184&0.000&0.000&0.000&-0.001&-0.254&0.176&-0.007&-0.280&0.205&0.001&0.000&0.029\\
      &slope&-0.035&-0.204&0.000&-0.077&-0.232&0.000&-0.044&-1.003&0.000&-0.146&-0.366&0.000&-0.179&-0.391&0.000&-0.914&-2.346&0.000\\
      &area&0.029&0.000&0.232&0.076&-0.017&0.322&0.000&0.000&0.001&0.173&0.000&0.525&0.193&-0.012&0.543&0.001&0.000&0.002\\
    &threshold&-0.014&-0.270&0.054&-0.056&-0.423&0.116&-0.001&0.000&0.000&-0.014&-0.342&0.212&-0.016&-0.315&0.256&0.002&-0.027&0.141\\
			\hline
    SBP&value&0.122&-0.017&0.475&0.110&-0.031&0.384&0.003&0.000&0.033&0.535&0.305&0.730&0.556&0.329&0.761&0.046&0.035&0.058\\
      &slope&0.000&-0.043&0.039&-0.009&-0.101&0.043&0.000&0.000&0.000&-0.006&-0.111&0.011&-0.015&-0.121&0.052&0.000&0.000&0.000\\
      &area&0.319&0.000&0.622&0.404&0.000&0.730&0.001&0.000&0.002&0.026&-0.118&0.397&0.052&-0.150&0.399&0.000&0.000&0.000\\
    &threshold&-0.568&-1.206&0.000&-0.609&-1.206&0.000&-0.286&-0.992&0.000&-0.174&-0.970&0.068&-0.345&-1.009&0.080&-0.097&-0.987&0.000\\
			\hline
    DBP&value&-0.023&-0.324&0.123&0.033&-0.196&0.321&0.000&0.000&0.000&-0.376&-0.630&0.000&-0.289&-0.582&0.006&-0.064&-0.095&0.027\\
      &slope&0.002&-0.032&0.063&0.059&-0.063&0.371&0.142&0.000&0.779&-0.002&-0.072&0.115&0.004&-0.137&0.156&0.000&0.000&0.000\\
      &area&-0.204&-0.432&0.000&-0.256&-0.535&0.000&-0.001&-0.002&0.000&-0.063&-0.406&0.022&-0.138&-0.484&0.040&0.000&0.000&0.000\\
      &threshold&0.120&-0.052&0.805&0.201&-0.129&0.857&0.099&0.000&0.776&0.979&0.000&1.888&0.707&-0.002&1.708&0.907&0.000&1.756\\
			\hline
    GLUCOSE&value&0.006&-0.094&0.131&0.004&-0.146&0.134&0.000&0.000&0.000&0.003&0.000&0.010&0.172&-0.016&0.385&0.001&0.000&0.009\\
      &slope&-0.059&-0.198&0.000&-0.024&-0.157&0.021&-0.313&-0.680&0.000&-0.006&-0.101&0.012&-0.003&-0.088&0.072&0.001&0.000&0.000\\
      &area&0.058&0.000&0.230&0.080&-0.015&0.276&0.000&0.000&0.000&0.157&0.000&0.423&0.134&-0.062&0.416&0.000&0.000&0.000\\
      &threshold&1.350&0.890&1.768&1.271&0.845&1.699&1.393&1.033&1.752&0.782&0.153&1.275&0.707&0.163&1.176&0.682&0.000&1.224\\
			\hline
    TOTCHL&value&0.059&-0.023&0.268&0.067&-0.059&0.268&0.005&0.000&0.035&0.041&-0.048&0.254&0.045&-0.156&0.263&0.018&0.000&0.043\\
      &slope&-0.022&-0.170&0.000&-0.112&-0.430&0.032&-0.006&0.000&0.000&-0.064&-0.319&0.000&-0.105&-0.346&0.027&-0.003&0.000&0.000\\
      &area&0.148&0.000&0.354&0.070&-0.054&0.286&0.001&0.000&0.002&0.131&0.000&0.341&0.119&-0.069&0.349&0.000&0.000&0.001\\
      &threshold&0.150&-0.038&0.751&0.213&-0.072&0.747&0.089&0.000&0.936&0.006&-0.288&0.353&0.004&-0.389&0.399&-0.006&-0.156&0.000\\
			\hline
		\end{tabular}
   \end{adjustbox}
	\end{table}

\end{document}